\documentclass[%
reprint,
superscriptaddress,
 amsmath,amssymb,
 aps,
prc
]{revtex4-2}
\usepackage{graphicx}           
\usepackage{dcolumn}            
\usepackage{bm}                 
\usepackage{hyperref}           
\usepackage[mathlines]{lineno}  
\usepackage{color}
\begin{document}
\preprint{APS/123-Nuclear Fission}
\title{A transport model description of Time-Dependent Generator Coordinate under Gaussian
    overlap approximation}
\author{Fangyuan Wang}
\affiliation{Department of Nuclear Physics, China Institute of Atomic Energy, Beijing 102413, People’s
Republic of China}%
\author{Yingxun Zhang}
\email{zhyx@ciae.ac.cn}
\affiliation{Department of Nuclear Physics, China Institute of Atomic Energy, Beijing 102413, People’s
Republic of China}%
\author{Zhipan Li}
\affiliation{School of Physical Science and Technology, Southwest University, Chongqing 400715, People’s
Republic of China}%
\date{\today}
%
\begin{abstract}
In this work, we derived a transport equation based on a generalized equation of time-dependent generator coordinate method (TDGCM) under the Gaussian overlap approximation (GOA). The transport equation is obtained by using quantum-mechanics phase space distributions under a ``quasi-particle" picture and strategy of Bogoliubov-Born-Green-Kirkood-Yvon (BBGKY) hierarchy. The theoretical advantage of this transport equation is that time evolution of $s$-body phase space density distribution is coupled with $s+1$-body phase space density distributions, and thus, non-adiabatic effects and dynamical fluctuations could be involved by more collective degrees and entanglement of phase space trajectories. 
In future, we will perform the numerical calculations for fission nuclei after obtaining collective inertia and potential energy surface (PES).
\end{abstract}
\maketitle
%
\section{Introduction}
After the discovery of nuclear fission by Hahn and Strassmann~\cite{hahn1939nachweis} in 1939, nuclear fission has become one of the most challenging topics in physics since it is a key ingredient for modeling nucleosynthesis~\cite{Baran2015}, energy production~\cite{Fermi1934}, medicine~\cite{Weber2020}, and nuclear safeguard~\cite{nichols2008}. Even with recent progress in experimental techniques, measurements of nuclear fission are not possible for all fission nuclei. Thus, theoretical simulations are mandatory for fully understanding the fission dynamics and complementing the missing data~\cite{bohr1939, Schunck2016, Andreyev17, Schmidt2018, Bender2020, Bulgac2020, Schunck2022}.

One kind of models is useful macroscopic model by taking into account shell effects, collective variables, and correlations between collective degree and single particles motions, such as Brownian shape motion ~\cite{Randrup2011prl, Randrup2011, Mumpower2020} and Langevin model~\cite{Ishizuka2017, Liu2019, Shimada2021}. With four or five collective degrees, it could depict fission dynamics process appropriately and reproduce fission yields distribution well.

Another kind of models is microscopic model, which based on nucleonic Hamiltonian and solve fission dynamics with Schr\"odinger or Dirac equation in time domain. For example, time-dependent density functional theory, such as time-dependent superfluid local density approximation (TDSLDA)~\cite{Bulgac2016, Bulgac2019}, Constrained and time-dependent Hartree-Fock calculations with dynamical Bardeen–Cooper–Schrieffer pairing correlations (CHF+BCS)~\cite{scamps2015, Scamps2018}, time dependent Hartree-Fock-Bogliubov (TDHFB)/TDHF+BCS~\cite{Scamps2012}, the adiabatic time-dependent HFB (ATDHFB)~\cite{Samuel2018} and time-dependent covariant density functional theory (TDCDFT)~\cite{Ren2022} describe fission process with full quantum microscopic approaches. At tremendous costs of computations, these models successfully describe the fission dynamics and predict the most probable fission yields. The great understanding of fission dynamics from microscopic model are obtained~\cite{Bulgac2016,Scamps2018}. However, describing fission yields distribution with these models is still a theoretical challenge due to the lacking of fluctuation in initial state and fission process. The efforts on this direction is to describe quantum fluctuation by a sampling of initial conditions followed by TDDFT\cite{Tanimura2017} but without quantum interference.

An alternative method to include the correlation is to represent a many-body wave function of the system with a mixture of states with different shapes. It stimulates the description of fission dynamics with time-dependent generator coordinate method under Gaussian overlap approximation (TDGCM+GOA)~\cite{Ring2004, Krappe12, Verriere2020}. In the TDGCM+GOA, fission is assumed as adiabatic process since the typical time for the motion of individual nucleons inside the fission nucleus (roughly $10^{-22}$ s) is roughly ten times smaller than the time scale of the system's collective deformation ($10^{-21}$ s)~\cite{Verriere2020}. Thus, the fission dynamics are approximately described in terms of a few shape coordinates.

Currently, most of the TDGCM+GOA calculations were performed by using only two degrees of 
freedom, usually $ \left(q_{20}, q_{30} \right) $ or $ \left( \beta_2, \beta_3 \right) $, under 
adiabatic assumption~\cite{Regnier2016, Regnier2016felix, Regnier2018felix}. While, the semiphenomenological and fully microscopic approaches illustrate that at least four or five collective variables play a role in the dynamics of fission~\cite{moller2001nuclear, Younes09, dubray2012numerical, schunck2014description,Bulgac2016, Eslamizadeh2018}. Regnier et al.~\cite{Regnier2017} have started some trials on the rigorous three degrees of freedom calculation of the PES for $^{240}\text{Pu} $ in the collective space $ \left( q_{20}, q_{30}, q_{40}\right) $, and their work is still in progress. In Ref.~\cite{Zhaojie21}, Zhao et al. did the calculations with the dynamical pairing degree of freedom as the third degree of freedom besides $\left(\beta_2, \beta_3 \right)$, and their results also demonstrate the importance of including more degree of freedom. Furthermore, one should note that an \textit{ad hoc} Gaussian smoothing have to be used at the end of the TDGCM+GOA calculations, to account for the fluctuations in particle number of the fragments due to both pairing effects and the finite number of particles in the neck region for points along the fission line. Thus, one would expect to develop a microscopic method that can include more collective degrees and the correlations to account for dynamical fluctuation and non-adiabatic effects, and reasonably describing fission dynamics and distribution of fission yields.

In this work, we derive a transport equation based on a generalized N-dimensional TDGCM+GOA equation to describe fission dynamics, in which non-adiabatic effects are introduced by more collective degrees, fluctuations are introduced by initial state fluctuation and entanglement of phase space trajectories. The paper is organized as follows: in Sec.\ref{theory}, the transport equation is obtained by using quantum-mechanics phase space distributions under a ``quasi-particle" picture and strategy of Bogoliubov-Born-Green-Kirkood-Yvon (BBGKY) hierarchy. One of the advantages of this hierarchy is that correlations from high-order degree can be involved in the evolution of the one-body phase-space density distribution. In Sec.\ref{numerical}, a numerical recipes for solving the transport equation is provided. Sec.\ref{summary} is the summary and outlook.

\section{Theory Framwork}\label{theory}

\subsection{Overview of TDGCM for fission}

For convenience, we briefly review the TDGCM +GOA theory which describes induced fission as a slow adiabatic process determined by a small number of collective degrees of freedom. Under the Griffin-Hill-Wheeler 
ansatz, the many-body state of fissioning system at any time reads
\begin{equation}
	\label{GCstate}
	|\Psi(t) \rangle=\int _{\mathbf{q}\in{E}}d\mathbf{q} |\phi(\mathbf{q})\rangle f(\mathbf{q},t).
\end{equation}
The set $\{|\phi(\mathbf{q})\rangle\}$ is a family of the generator states which are the 
solutions of a constrained Hartree-Fock-Bogoliubov equation. $f(\mathbf{q},t)$ is the 
complex-valued weights of the quantum mixture of states. The generator coordinate 
$\mathbf{q}=\{q_1, ..., q_N\}$, and each of these $q_i$ is a collective variable chosen 
based on the physics of fission. 

The time-dependent Schr\"odinger equation for the many-body state of fission system $|\Psi(t)\rangle$,
\begin{equation}
	\label{TLSeq}
	(\hat{H}-i\hbar \frac{d}{dt})|\Psi(t)\rangle=0,
\end{equation}
can yield an equation of the unknown weight function $f(\mathbf{q},t)$, i.e., the 
Hill-Wheeler equation with time-dependent form,
\begin{equation} \label{TDHW}
	\int \text{d} \mathbf{q}' \langle \phi_{\mathbf{q}}|\left( \hat{H}-i\hbar \frac{d}{dt}\right) |\phi_{\mathbf{q}'}\rangle f(\mathbf{q}',t)=0.
\end{equation}
Here, $\hat{H}$ is the Hamiltonian acting on the full many-body system. Principally, 
Eq.~(\ref{TDHW}) can be solved numerically, but it needs a tremendous amount of 
computations. To overcome these difficulties, a popular approach named as 
Gaussian overlap approximation (GOA) is used. The simplest formulation of GOA assumes that the overlap between two generator states $\langle\phi_{\mathbf{q}}|\phi_{\mathbf{q}'}\rangle$ 
has a Gaussian shape,
\begin{equation}
	\mathcal N(\mathbf q,\mathbf q')=\langle\phi_{\mathbf{q}}|\phi_{\mathbf{q}'}\rangle\equiv \exp\left[-\frac{1}{2}(\mathbf{q}-\mathbf{q}')^t \mathbf{G}(\bar{\mathbf{q}})(\mathbf{q}-\mathbf{q}')\right].
\end{equation}
$\mathcal N(\mathbf q,\mathbf q')=\langle\phi_{\mathbf{q}}|\phi_{\mathbf{q}'}\rangle$ is peaked functions for $\mathbf{q}=\mathbf{q}'$, and $\mathbf{\bar{q}}=(\mathbf{q}+\mathbf{q}')/2$. By changing a new collective coordinate $\boldsymbol{\alpha}$ by the relation
\begin{equation}
	\boldsymbol{\alpha}(\mathbf{q})=\int_{a\in C_{0}^{q}} G^{1/2}(\mathbf{a}) d\mathbf{a},
\end{equation}
in terms of which the overlap matrix becomes
\begin{equation}
	\mathcal N(\boldsymbol{\alpha},\boldsymbol{\alpha}')=\exp\left[-\frac{1}{2}(\boldsymbol{\alpha}-\boldsymbol{\alpha}')^2\right],
\end{equation}
$\mathbf{G} \left( \mathbf{q} \right) $ is the metric of new coordinates of $\mathbf{ \alpha( \mathbf{ q})}$, and $G\left( \mathbf{q} \right)$ is the determinant of $\mathbf{G}$.

Within this approximation, the time-dependent Hill-Wheeler equation is reduced to a local, time-dependent 
Schr\"odinger-like equation as
\begin{equation} \label{TD-SLQ}
	i\hbar\frac{\partial g(\mathbf{q},t)}{\partial t}=\hat{H}_{coll}(\mathbf{q})g(\mathbf{q},t).
\end{equation}
$g(\mathbf{q},t)$ is related to the weight function $f(\mathbf{q},t)$ as $g=\mathcal N^{1/2} f$, and contains all the information about the fission dynamics of system~\cite{Verriere2020}. The collective Hamiltonian $\hat{ H }_{ coll }(\mathbf{q})$ is a local operator acting on $g(\mathbf{q},t)$,
\begin{eqnarray}
\label{Hcoll}
&& \hat{H}_{coll}(\mathbf{q}) =\\\nonumber
&& \left[ -\frac{\hbar^2}{2} \sum_{kl}{\frac{1}{\sqrt{ G \left(\mathbf{q} \right) }} \frac{\partial}{\partial q_k} \sqrt{  G \left(\mathbf{q} \right)} B_{kl} \left( \mathbf{q} \right)\frac{\partial}{\partial q_l} + V \left( \mathbf{q} \right) }
	\right].
\end{eqnarray}
The potential $V(\mathbf{q})$, 
\begin{equation}
	V(\mathbf{q})=\langle \mathbf{q}|\hat{H}|\mathbf{q}\rangle-\epsilon_{0}(\mathbf{q}),
\end{equation}
with the zero-point energy-correction
\begin{equation}
	\epsilon_{0}=\frac{1}{2}G^{ij}(\mathbf{q})\frac{\partial^2 h}{\partial q^i\partial q'^j} \bigg|_{\mathbf{q}=\mathbf{q}'}.
\end{equation}
The symmetric collective inertial tensor $\mathbf{B(q)}\equiv B_{ij}(\mathbf{q})$,
\begin{multline}
B_{kl}(\mathbf{q})=\frac{1}{2\hbar^2} G^{km}(\mathbf{q}) \bigg( \frac{\partial^2 h(\mathbf{q},\mathbf{q'})}{\partial q^m\partial q'^{n}} -\frac{\partial^2 h(\mathbf{q},\mathbf{q'})}{\partial q^m\partial q^{n}} \\
+ \left\lbrace \begin{array}{c} i\\ mn\end{array} \right\rbrace\frac{\partial h(\mathbf{q},\mathbf{q'})}{\partial q^i} \bigg) \bigg|_{ \mathbf{ q } = \mathbf {q }'} G^{ nl }( \mathbf{q} ),
\end{multline}
the expression in braces is the Christoffel symbol of the second kind. $h(\mathbf q, 
\mathbf{q}')$ is
\begin{equation}
	h(\mathbf q,\mathbf{q}')=\frac{\left<\phi_\mathbf{q}|\hat H|\phi_{\mathbf{q}'}\right>}{\left<\phi_\mathbf{q}|\phi_{\mathbf{q}'}\right>}.
\end{equation}
They are usually calculated from the nuclear Hamiltonian $\hat{H}$ and the generator 
states $|\phi_{\mathbf{q}}\rangle$ with HFB~\cite{Regnier2016} or RMF+BCS~\cite{Tao2017}. 

The number of collective degree of freedom are usually selected as $N=2$, and shape coordinates $\mathbf{q}$ are the multipole moments $Q_{20}$ 
and $Q_{30}$ in Ref.~\cite{Regnier2016, Regnier2016felix, Regnier2018felix}, or $\beta_2$ and $\beta_3$ as in Refs.~\cite{Tao2017, Zhaojie21}, and $ G \left(\mathbf{ q } \right) = 1 $ Ref.~\cite{Regnier2016felix}. In this case, the equation of TDGCM+GOA is 
\begin{eqnarray}\label{tdgcm2-simp}
	&&i \hbar \frac{\partial}{\partial t}g\left( q_1, q_2;
	t \right) =\\\nonumber
	&& \left[ -\frac{\hbar^2}{2} \sum_{kl}{ \frac{\partial}{\partial
			q_k}  B_{kl} \left( q_1, q_2 \right) \frac{\partial}{\partial q_l} +
		V\left( q_1, q_2 \right) }  \right] g \left(q_1, q_2; t
	\right). 
\end{eqnarray}
This equation has been solved by the software package \verb*|FELIX-1.0|~\cite{Regnier2016felix} or \verb*|FELIX-2.0|~\cite{Regnier2018felix} with finite element method.

\subsection{A transport equation for N-dimensional TDGCM+GOA}

The TDGCM has achieved great progress on describing the fission dynamics~\cite{Regnier2016, Regnier2016felix, Tao2017, Regnier2018felix, Zhao2020, Zhaojie21}, but previous calculations in Refs.~\cite{moller2001nuclear, Younes09, dubray2012numerical, schunck2014description, Regnier2017} also showed it is necessary to include more degrees to describe nonadiabatic effects which may arise from the coupling between collective and intrinsic degrees of freedom, and invovle dynamical fluctuations to describe fission products distributions. Now, the question is that can we effectively involve more collective degrees of freedom into the equation with two degrees that we are currently using?

Principally, nuclear shape can be described by an expansion in spherical harmonics, i.e.,
\begin{equation}
	R(\theta,\phi,t)=R_0\left (1+\sum_{\lambda=0}^{N}\sum_{\mu=-\lambda}^{\lambda}\alpha_{\lambda\mu}^\ast(t)Y_{\lambda\mu}(\theta,\phi)\right ).
\end{equation}
The number of shape coordinates (or collective degree of freedom) $N$ depends on the choice of collective coordinates or generator coordinates and stage of fission. In the stage of fissionning system from the quasistationary initial state to the outer fission barrier, evolution is slow and fission process can be described by a small number of collective degree, i.e., a small $N$, with adiabatic approximation\cite{Negele1978}. In the stage of saddle-to-scission, the nucleus quickly elongates toward scission and non-adiabatic effects have to be considered and $N$ may vary with the stage of fission process. Thus, the collective wave function is presented with $N$ degrees, i.e., $g(q_1, q_2, ..., q_N)$, for fissioning system, and Eq.(\ref{TD-SLQ}) will become a generalized N-dimensional TDGCM+GOA equation since the degrees are not limited to a few.

In this work, we interpret the wave function of $g(q_1, q_2, ..., q_N)$ for fissioning system as a wave function for `N-body quasiparticles in 1-Dimension space' system (NB1D), and a convention $g_N(q_1, q_2, ..., q_N)$ is used in following to represent the wave function with particle from 1 to N. Then, we derive a transport equation  which can effectively couple one more collective degree of freedom to the degrees currently used. 
Firstly, we perform Wigner transformation~\cite{wigner1932quantum} on NB1D wave function $ g_N \left( q_1, \cdots, q_N \right) $ to get their quantum mechanically phase space density $f_N$ as,
\begin{eqnarray}\label{eq003}
&& f_N\left( q_1,\cdots, q_N; p_1,\cdots, p_N \right) \\\nonumber
&=& \frac{1} {\left( \pi \hbar \right)^N} \int \cdots \int
    dy_1\cdots dy_N g_N^* (q_1-y_1, \cdots, q_N-y_N )\\\nonumber
&&g_N(q_1+y_1, \cdots , q_N+y_N ) \\\nonumber
&&\times \exp [-2i(p_1\cdot y_1+\cdots + p_N\cdot y_N) / \hbar ]. 
\end{eqnarray}
Here $ p_i $ is the conjugated momentum of $q_i $ for quasi-particle $i$. After a trivial deviation, the corresponding transport equation reads,
\begin{eqnarray}\label{eq-fnt-final}
	\frac{\partial f_N}{\partial t} &=& -\sum_{kl}  \bar{\mathcal {B}}_{kl}^{(b)} ( \mathbf{q} ) p_k\frac{\partial f_N}{\partial q_l}  \\\nonumber
	&&+\sum\limits_{\lambda=1,3,\cdots}(\frac{\hbar}{2i})^{\lambda_1 + \cdots + \lambda_N-1} \frac{1}{\lambda_1 ! \cdots \lambda_N !} \\\nonumber
&&\times\frac{ 
		\partial^{\lambda_1 + \cdots + \lambda_N} V_N \left( \mathbf{q} \right) } {\partial 
		q_1^{\lambda_1} \cdots \partial q_N^{\lambda_N}} 
	\frac{ \partial^{\lambda_1 + \cdots + \lambda_N} f_N  } {\partial 
		p_1^{\lambda_1} \cdots \partial p_N^{\lambda_N}}\\\nonumber
&=& -\sum_{kl}  \bar{\mathcal {B}}_{kl}^{(b)} ( \mathbf{q} ) p_k\frac{\partial f_N}{\partial q_l}+\sum_{l}\frac{\partial V_N}{\partial q_l}\frac{\partial f_N}{\partial p_l}\\\nonumber
	&&+\sum\limits_{\lambda=3,\cdots}(\frac{\hbar}{2i})^{\lambda_1 + \cdots + \lambda_N-1} \frac{1}{\lambda_1 ! \cdots \lambda_N !} \\\nonumber
&&\times\frac{ 
		\partial^{\lambda_1 + \cdots + \lambda_N} V_N \left( \mathbf{q} \right) } {\partial 
		q_1^{\lambda_1} \cdots \partial q_N^{\lambda_N}} 
	\frac{ \partial^{\lambda_1 + \cdots + \lambda_N} f_N  } {\partial 
		p_1^{\lambda_1} \cdots \partial p_N^{\lambda_N}}.		
\end{eqnarray}
Here, $\lambda=\sum_{i=1}^N \lambda_i$ and $\bar{\mathcal {B}}_{kl}^{(b)} ( \mathbf{q} )$ is an effective collective inertia, which is defined as 
\begin{equation}
\bar{\mathcal{B}}_{kl}^{(b)}(\mathbf{q}) = \bigg( B_{kl}(\mathbf{q-y^*_{(1b)}})+B_{kl}(\mathbf{q-y^*_{(2b)}}) \bigg )/2.
\end{equation}
$\mathbf{y}^*_{(1b)}$ and $\mathbf{y}^*_{(2b)}$ are corrections on $\mathbf{q}$, and their origins can be found in appendix~\ref{app-fnt}. $V_N(\mathbf{q})$ is N-body potential. 

Principally, $V_N(\mathbf{q})=V(q_1, q_2, \cdots, q_N)$. If the N-body potential is calculated from the two-body interaction, i.e.,
\begin{equation}
	V(q_1, q_2, \cdots, q_N)=\sum_{i\le j} V(q_i,q_j), 
\end{equation}
the transport equation is simplified as,
\begin{eqnarray}\label{eq-fnt-final2b}
	\frac{\partial f_N}{\partial t} &=& -\sum_{kl}  \bar{\mathcal {B}}_{kl}^{(b)} ( \mathbf{q} ) p_k\frac{\partial f_N}{\partial q_l} + \frac{1}{2}\sum_{ k, m \neq k } \frac{\partial V_{km}}{\partial q_k} \frac{\partial f_N } { \partial p_k } .
\end{eqnarray}
$V_{km}$ is two-body potential between quasi-particle $k$ and $m$, which can be obtained by HFB/RMF+BCS as in Refs.\cite{Regnier2016,Tao2017}. When the N-body potential is obtained from multi-dimensional PES by HFB/RMF+BCS, transport equation of Eq.(\ref{eq-fnt-final}) should be used.

A standard procedure to solve the $N$-body transport equation is to use BBGKY hierarchy, in which one-body degrees of freedom (DOF) is coupled to two-body DOF that are themselves coupled to three-body DOFs and so forth. As an example, we present the time evolution of $f_s$ under the condition of $V(q_1, q_2, \cdots, q_N)=\sum_{i\le j} V(q_i,q_j)$. The $s$-body phase space density distribution $f_s$ is defined as,
\begin{eqnarray}\label{fs}
&&	f_s(q_1, \cdots, q_s, p_1, \cdots, p_s)\\\nonumber
&&=\frac{1}{\Omega^{N-s}}\int f_N(q_1, \cdots, q_N, p_1, \cdots, p_N)d\Gamma_{s+1}\cdots d\Gamma_N,\\\nonumber
&&	 d\Gamma_i=dq_idp_i,
\end{eqnarray}
here, $\Omega$ is volume in phase space. Thus,
\begin{eqnarray}
&&\frac{\partial f_s(q_1, \cdots, q_s, p_1, \cdots, p_s)}{\partial t}\\\nonumber
&=&\frac{1}{\Omega^{N-s}}\int \frac{\partial f_N(q_1, \cdots, q_N, p_1, \cdots, p_N)}{\partial t}d\Gamma_{s+1}\cdots d\Gamma_N\\\nonumber
&=&\frac{1}{\Omega^{N-s}}\int\bigg[-\sum_{kl}  \bar{\mathcal {B}}_{kl}^{(b)} ( \mathbf{q} ) p_k\frac{\partial f_N}{\partial q_l}  + \sum_{ 1\le k< m \le N } \frac{\partial V_{km}}{\partial q_k} \frac{\partial f_N } { \partial p_k } \bigg]\\\nonumber
&& d\Gamma_{s+1}\cdots d\Gamma_N.
\end{eqnarray}

One should note that the derivation in this case is different than
a system with fixed-mass many particles, because the inertial $\bar{\mathcal {B}}_{kl}^{(b)} ( \mathbf{q} )$ depends on collective coordinates. To overcome this difficulty, we move the $\bar{\mathcal {B}}_{kl}^{(b)} ( \mathbf{q} )$ out from the integration by assuming the following relationship, i.e., 
\begin{eqnarray}\label{Int-Approx}
	&&\int B_{kl}(\mathbf{q}) O(\mathbf{q},\mathbf{p})d\Gamma_{s+1}\cdots d\Gamma_{N}=\\\nonumber
	&&B_{kl}(\mathbf{q_s}, q_{s+1}^*,\cdots, q_{N}^*)\int  O(\mathbf{q},\mathbf{p})d\Gamma_{s+1}\cdots d\Gamma_{N},
\end{eqnarray}
The $q^{*}_{s+1}, \cdots, q^*_{N}$ depend on $O(\mathbf{q},\mathbf{p})$, and its values will be fixed once the $O(\mathbf{q},\mathbf{p})$ is determined.

By using above relationship, we get
\begin{eqnarray}\label{fst-final}
	\frac{\partial f_s}{\partial t}	&=&-\sum_{k=1}^s\sum_{l=1}^s \bar{\mathcal {B}}_{kl}^{(b)} (\mathbf{q_s},q_{s+1}^*,\dots, q_{N}^*) p_k\frac{\partial f_s}{\partial q_l}\\\nonumber
	&-&\sum_{k=s+1}^N\sum_{l=1}^s \bar{\mathcal {B}}_{kl}^{(b)} (\mathbf{q_s},q_{s+1}^*,\dots, q_{N}^*) p_k^*\frac{\partial f_s}{\partial q_l}\\\nonumber
	&+&\sum_{1\le k< m\le s } \frac{\partial V_{km}}{\partial q_k} \frac{\partial f_s } { \partial p_k }+\frac{N-s}{\Omega}\int \sum_{k=1}^s \left(\frac{\partial V_{k,s+1}}{\partial q_k}\right)\\\nonumber
	&\times&\left(\frac{\partial f_{s+1}}{\partial p_k}\right)d\Gamma_{s+1}.\\\nonumber
\end{eqnarray}
The collective inertial $\bar{\mathcal {B}}_{kl}^{(b)}  (\mathbf{q_s},q_{s+1}^*,\dots, q_{N}^*)$ only varies with $\mathbf{q_s}$, since the values of $q_{s+1}^*,\dots, q_{N}^*$ will be fixed once the integrand was selected. But the values of $\bar{\mathcal {B}}_{kl}^{(b)}  (\mathbf{q_s},q_{s+1}^*,\dots, q_{N}^*)$ could be different from $B_{kl} (\mathbf{q_s})$. The details of the derivation are in appendix~\ref{app-fst}. 

By expressing the fourth terms on the right side of Eq.~(\ref{fst-final}) as $\delta I_{coll}$, the transport equation can be rewritten as,
\begin{eqnarray}\label{fst-final-Icoll}
	\frac{\partial f_s}{\partial t}	&+&\sum_{k=1}^s\sum_{l=1}^s \bar{\mathcal {B}}_{kl}^{(b)}  (\mathbf{q_s},q_{s+1}^*,\dots, q_{N}^*) p_k\frac{\partial f_s}{\partial q_l}\\\nonumber
	&+&\sum_{k=s+1}^N\sum_{l=1}^s \bar{\mathcal {B}}_{kl}^{(b)}  (\mathbf{q_s},q_{s+1}^*,\dots, q_{N}^*) p_k^*\frac{\partial f_s}{\partial q_l}\\\nonumber
	&-&\sum_{1\le k < m\le s } \frac{\partial V_{km}}{\partial q_k} \frac{\partial f_s } { \partial p_k }=\delta I_{coll}.\\\nonumber
\end{eqnarray}
with
\begin{eqnarray}\label{DR}
	\delta  I_{coll}=\frac{N-s}{\Omega}\int \sum_{k=1}^s \left(\frac{\partial V_{k,s+1}}{\partial q_k}\right)\left(\frac{\partial f_{s+1}}{\partial p_k}\right)d\Gamma_{s+1}.
\end{eqnarray}
As one can see that the $\delta I_{coll}$ is related to the phase space density $f_{s+1}$, and the potential between $k$ and $s+1$, i.e., $V_{k,s+1}$.

\section{Numerical recipe for solving transport equation}\label{numerical}

In this section, we discuss practical numerical recipe of the time evolution of $f_2$. In the following discussions, we take $q_1=\beta_2$ and $q_2=\beta_3$.

Given $s=2$, the time evolution of $f_2$ becomes,
\begin{eqnarray}\label{f2t}
	&& \frac{ \partial f_2 (q_1,q_2;p_1, p_2) }{ \partial t } \\\nonumber
	&+& \sum_{k=1}^{2} \sum_{ l = 1 }^{ 2 }
	p_k \bar{\mathcal {B}}_{kl}^{(b)} (q_1,q_2,q_3^*,\cdots, q_N^*) \frac{\partial f_2}{\partial q_l}\\\nonumber
	&+&\sum_{k=3}^{N} \sum_{ l = 1 }^{ 2 }
	p_k^* \bar{\mathcal {B}}_{kl}^{(b)} (q_1,q_2,q_3^*,\cdots, q_N^*) \frac{\partial f_2}{\partial q_l} \\\nonumber
	&-&\sum_{k=1}^{2} \sum_{ l \neq k }^{ 2 } \frac{1}{2}\frac{\partial V_{kl}}{\partial q_k}  \frac{ \partial f_2 }{ \partial p_k } 
	=\delta I_{coll}.
\end{eqnarray}
The time evolution of $f_2$ is not only related to the collective inertia $\mathcal{B}_{kl}$ and potential $V_{kl}$, but also related to $f_3$ and potential between $q_{i=1,2}$ and $q_3$ which actually reflect the high order correlation between different shape coordinates.

\subsection{Initialization}

For the TDGCM+GOA equation, the starting point is a collective wave packet at initial 
time, which represents the compound nucleus after excitation by absorption of a low-energy
neutron or photon. One choice is to use the quasibound state~\cite{Regnier2016}, i.e., collective ground state $g_0(\mathbf{q},t=0)$, and $\mathbf{q}=\{q_1, q_2\}$. Its modulus is roughly a Gaussian centered on the minimum of the potential $V_{min}(\mathbf{q})$, which is achieved by extrapolating inner potential barrier with a quadratic form. The width of this Gaussian is characterized by a width close to the dimension of the first potential well. To describe fission, $g_0$ has to boost in $q_1(\beta_2)$ direction for simulating the fission events, i.e.,
\begin{equation}\label{initi-gs}
	g(\mathbf{q},t=0)=g_0(\mathbf{q})\exp(ik q_1), 
\end{equation}
since its average energy is below the fission barrier. The amplitude $k$ of the boost is 
determined so that the average energy of the initial state lies few MeV above the inner 
fission barrier.

For the transport equation described in this work, we need to do the initialization in 
$\{q_1, q_2; p_1, p_2\}$ space according the phase space density $f_2$. The initial 
$f_2$, which represents the compound nucleus after the absorption of a low-energy neutron,
can be realized by doing Wigner transformation on $g_0(\mathbf{q})$, i.e.,
\begin{eqnarray}
	&& f_2(q_1,q_2;p_1, p_2,t=0)=\\\nonumber
	&& \frac{1}{(\pi\hbar)^2}\int dy_1dy_2 g_0^*(\mathbf{q+y})g_0(\mathbf{q-y})e^{2i\mathbf{p}\cdot\mathbf{y}/\hbar}.
\end{eqnarray}

Numerically, a test particle method is used to describe $f_2(q_1,q_2;p_1, p_2,t=0)$, which means each quasiparticle is replaced by a large number of test particles and the method was first proposed by Wong in nuclear Vlasov model~\cite{wong1982dynamics}.
\begin{eqnarray}
&&	f_2(q_1,q_2;p_1, p_2,t=0)=\\\nonumber
&&	\frac{1}{N_{test}}\sum_{k=1}^2\sum_{i=1}^{N_{test}} \delta({q}_{ki}-{\bar{q}}_{ki}(t))\delta({p}_{ki}-{\bar{p}}_{ki}(t)).
\end{eqnarray}
${q}_{ki}$ and ${p}_{ki}$ are the time-dependent coordinates and momenta of the test-particle $i$ for particle $k$=1 or 2. ${\bar{q}}_{ki}(t=0)$ and ${\bar{p}}_{ki}(t=0)$ are sampled according to the $f_2$. $N_{test}$ is the number of test particles. Once ${\bar{p}}_{ki}(t=0)$'s are obtained, $\bar{p}_{1i}(t=0)$ is boosted as $\bar{p}_{1i}(t=0)+k$. $k$ can be obtained as same as in TDGCM initialization.

\subsection{Time evolution}

%
To solve the transport equations with test particle method, we separately treat the left and right hand of Eq.(\ref{f2t}) as same as in the transport model used for simulating the heavy ion collisions\cite{WOLTER2022103962}.

The equation of motion of test particles under the mean field can be obtained by comparing,
\begin{equation}\label{fs-trans}
\frac{df_2}{dt}=\frac{\partial f_2(q_1,q_2;p_1,p_2)}{\partial t}+\sum_{k=1}^2 \left[ \frac{\partial f_2}{\partial q_k}\frac{\partial q_k}{\partial t}+\frac{\partial f_2}{\partial p_k}\frac{\partial p_k}{\partial t}\right]=0,
\end{equation}
to Eq.(\ref{f2t}) without considering the collision term, i.e.,
\begin{eqnarray}\label{fst-final-vlasov}
	\frac{\partial f_2}{\partial t}&=& -\sum_{k=1}^2\sum_{l=1}^2 \bar{\mathcal {B}}_{kl}^{(b)}  (q_1,q_2,q_{3}^*,\dots, q_{N}^*) p_k\frac{\partial f_2}{\partial q_l}\\\nonumber
	&-&\sum_{k=3}^N\sum_{l=1}^2 \bar{\mathcal {B}}_{kl}^{(b)}  (q_1,q_2,q_{3}^*,\dots, q_{N}^*) p_k^*\frac{\partial f_2}{\partial q_l}\\\nonumber
	&+&\sum_{1\le k < m\le 2 } \frac{\partial V_{km}}{\partial q_k} \frac{\partial f_2 } { \partial p_k }.
\end{eqnarray}
The equation of motion of test particle $ki$ becomes,
\begin{eqnarray}
	\dot{q}_{ki} &=& \sum_{l=1}^2 \bar{\mathcal {B}}_{kl}^{(b)}  (q_1,q_2,q_{3}^*,\dots, q_{N}^*) p_{li}\label{qdot}\\\nonumber 
	&+&\sum_{l=3}^N \bar{\mathcal {B}}_{kl}^{(b)}  (q_1,q_2,q_{3}^*,\dots, q_{N}^*) p^*_{li}\\
	\dot{p}_{ki} &=& -\frac{1}{2}\sum_{1\le l\le 2 } \frac{\partial V_{kl}}{\partial q_{ki}}\label{pdot},
\end{eqnarray}
The abbreviation of ${ki}$ in the lower index means particle $k$ and its $i$th test-particle, i.e., $k=1,2$ and $i=1, \cdots, N_{test}$. 

As shown in Eq.(\ref{qdot}), the evolution of position of test particle $ki$ not only depend on the collective inertia $\bar{\mathcal {B}}_{kl}^{(b)}  (q_1, q_2,q_{3}^*,\dots, q_{N}^*)$ and momentum of $p_{li}$ with $l=1$ and 2, but also on collective inertia $\bar{\mathcal {B}}_{kl}^{(b)}  (q_1, q_2,q_{3}^*,\dots, q_{N}^*)$ and effective momentum of particle of $p_{ki}^*$ with $l\ge 3$. In the calculations, one can approximate $\bar{\mathcal {B}}_{kl}^{(b)}  (q_1, q_2,q_{3}^*,\dots, q_{N}^*)=\eta(q_{3}^*,\dots, q_{N}^*) B_{kl} (q_1,q_2)$ for $l\le 2$, in which the parameter $\eta$ depend on the selection of $q_{k\ge 3}$. Alternatively, one can use $\eta$ as a phenomenological parameter to fit the data of fission yield. The value of $\bar{\mathcal {B}}_{kl}^{(b)}  (q_1, q_2,q_{3}^*,\dots, q_{N}^*)=\eta(q_{3}^*,\dots, q_{N}^*) B_{kl} (q_1,q_2)$ for $l\ge 3$ can be learned if the three-dimensional collective inertia can be provided. Within the framework of this equation, the correlations beyond $q_1$ and $q_2$ are involved via  $\bar{\mathcal {B}}_{kl}^{(b)}  (q_1, q_2,q_{3}^*,\dots, q_{N}^*)$. In the second term of Eq.(\ref{qdot}), the 
contribution of $p^*_{ki}$ can also be thought as friction effects. 

The momentum of test particle update according to Eq.(\ref{pdot}), and it will depend on the potential $V_{km}$. Inside the scission line (hypersurface), potential $V_{km}$ can be obtained by RMF+BCS/HFB model. Out of the scission hypersurface, the fissioning trajectories will not go back and one can set the potential $\partial V_{km}/\partial q_k=0$. 

For high order correlation term in Eq.~(\ref{DR}), i.e., the collision term in our approach, it comes from,
\begin{eqnarray}\label{Coll-Num}
	\delta  I_{coll}&=\frac{N-2}{\Omega}\int \sum_{k=1}^2 \left(\frac{\partial V_{k,3}}{\partial q_k}\right)\left(\frac{\partial f_{3}}{\partial p_k}\right)d\Gamma_{3}.
\end{eqnarray}
Suppose $f_3$ can be expressed as,
\begin{equation}
    f_3(q_1, q_2, q_3; p_1, p_2, p_3)=f_2(q_1,q_2;p_1,p_2)f_1(q_3;p_3),
\end{equation}
Thus, 
\begin{equation}\label{coll3}
    \delta I_{coll}=\sum_{k=1}^2 \left(\frac{\partial \bar{\Phi}_k}{\partial q_k}\right)\left(\frac{\partial f_{2}}{\partial p_k}\right),
\end{equation}
and potential $\bar{\Phi}_k$ means,
\begin{equation}\label{phi3}
    \bar{\Phi}_k=\frac{N-2}{\Omega}\int V_{k,3}(q_k,q_3) f_{1}(q_3,p_3) d\Gamma_{3},
\end{equation}
which reflect the potential of $k$ particle felt by surrounding particles. However, exact calculations of Eq.(\ref{coll3}) and Eq.(\ref{phi3}) are impossible since they always beyond one more degree we have.

One effective way to handle the collision term is to use a random collision among test particles. 
For example, one first select three particles among $2N_{test}$ test particles according to ``collision section'', and then, perform a random collision as follows,
\begin{equation}
 p_{k1}+p_{k2}+p_{k3}=p'_{k1}+p'_{k2}+p'_{k3}.  
\end{equation}
$p'_{k1}$, $p'_{k2}$ and $p'_{k3}$ will be determined by using the Monte-Carlo sampling under the momentum conservation,
\begin{eqnarray}
    p'_{k1}&=&(p_{k1}+p_{k2}+p_{k3})*\xi_1,\\
    (p'_{k2}+p'_{k3})&=&(p_{k1}+p_{k2}+p_{k3})*(1-\xi_1),\\
    p'_{k2}&=&(p'_{k2}+p'_{k3})*\xi_2,\\
    p'_{k3}&=&(p'_{k2}+p'_{k3})*(1-\xi_2).
\end{eqnarray}
$\xi_1$ and $\xi_2$ are the random number satisfy a certain distribution. The collision among test particles during the evolution describe entanglement among different trajectories of test particles. In practical calculations, the collision probability can be adjusted by introducing a `cross section' of three-body collision. After random collision, the values of momentum of test particle will be randomly modified. As a result, the fluctuation on $\mathbf{q}$ will be automatically involved and the mass distribution of fission fragment can be expected.

\subsection{Fission fragments distributions}


In this work, we only focused on fission fragment mass/charge distribution. First, one need to search scission line (or hypersurface) on potential energy surface, which is composed from many scission points $\mathbf{q}_{sci}$. Inside the scission line(hypersurface), the nucleus is whole. Out of scission line (hypersurface), the system becomes two well-separated fragments which are connected by a thin neck. Thus, each scission points $\mathbf{q}_{sci}$ is associated with a given fragmentation $(A_L, A_R)$. $A_L$ and $A_R$ 
means the mass of fission fragment in the left and right of neck, respectively. The fission fragment mass can be obtained from the integration of single-body density over the
domain of left or right of neck,
\begin{eqnarray}
	A_L&=&\int_{\mathbf r\in L} d\mathbf{r}\rho(\mathbf r), \\\nonumber
	A_R&=&\int_{\mathbf r\in R} d\mathbf{r}\rho(\mathbf r). \\\nonumber
\end{eqnarray}
Here, $R$ and $L$ means the region of right and left of neck of fissioning system. 
$\rho(\mathbf r)$ is constructed from the collective coordinates.

In transport model approach, the probability of measured the fission fragment $A_L$ and 
$A_R$, i.e., $Y(A_L)$ and $Y(A_R)$ can also be obtained from the time integrated flux 
through the hypersurface element $\mathbb{S}$. By using the test particle method, it will be 
obtained by counting the number of test particles across the hypersurface $\mathbb S$ at 
scission point, i.e., with $ t \rightarrow + \infty $,
\begin{eqnarray}\label{YAxi}
Y(A,\mathbb S) &=& \int\int_{\mathbf{q}_s>\mathbf{q}_{sci}} f_s(\mathbf{q}_s,\mathbf{p}_s,t) d\mathbf{p}_sd\mathbf{q}_s\\\nonumber
&=&\frac{1}{2N_{test}}\sum_{k=1}^2\sum_{i=1}^{N_{test}} \Theta(q_{ki}-\bar{q}_{ki,sci}(t)).
\end{eqnarray}
Thus, the yield of mass of fragment
\begin{equation}
	Y(A)=\sum_{\mathbb S} Y(A,\mathbb S).
\end{equation}
The sum on $\mathbb S$ runs over the whole scission hypersurface.

\section{Summary and Outlook}\label{summary}
In this work, we derive a transport equation based on a generalized N-dimensional TDGCM+GOA equation to describe fission dynamics. The transport equation is obtained by using quantum-mechanics phase space distributions under a ``quasi-particle" picture and strategy of BBGKY hierarchy. The advantages of this transport equation is that time evolution of $s$-body phase space density distribution is coupled with $s+1$-body phase space density distributions. Thus, one can expect that non-adiabatic effects and dynamical fluctuations could be introduced by involving more collective degrees and entanglement of phase space trajectories. 

Different than directly solving the TDGCM+GOA equation with finite elements method, our 
approach is realized by using the test particle method. The coordinates and momentum of test particles in initialization of fissioning system are sampled according to initial phase space density distribution of system. The time evolution of test particles are governed by a Hamiltonian-like equation coupled with a random scattering between test particles, which will naturally provide a fluctuation on the mass of fission fragments. Finally, the numerical results on this transport equations are still on the way, since we need to select reasonable collective coordinates to obtain the results of PES and collective inertia, and the high order effects of degree of freedom on collective inertia should also be investigated in this approach before presenting numerical results. 

\section*{Acknowledgements}
The authors thank Zhuxia Li, Zaochun Gao and Siyu Zhuo for reading the manuscript and providing useful feedback. This work was partly supported by the National Natural Science Foundation of China Nos. 11875323, 12275359, 11875225, 11705163, 11790320, 11790323, and 11961141003, the National Key R\&D Program of China under Grant No. 2018 YFA0404404, the Continuous Basic Scientific Research Project (No. WDJC-2019-13, BJ20002501), Key Laboratory of Nuclear Data foundation (No.JCKY2022201C158) and the funding of China Institute of Atomic Energy YZ222407001301. The Leading Innovation Project of the CNNC under Grant No. LC192209000701, No.  LC202309000201.

\appendix
\section{Derivation of transport equation}\label{app-fnt}

The equation of motion of $g_N$ is determined by the Schr\"odinger-like equation, i.e., 
\begin{eqnarray}\label{eq-g-append}
	&&i \hbar \frac{\partial}{\partial t}g_N\left( \mathbf{q},t \right) =\\\nonumber
	&& \left[ -\frac{\hbar^2}{2} \sum_{kl}{ \frac{\partial}{\partial
			q_k}  B_{kl} \left(\mathbf{q} \right) \frac{\partial}{\partial q_l} +
		V_N \left( \mathbf{q} \right) }  \right] g_N \left( \mathbf{q}, t
	\right). 
\end{eqnarray}
By using the Wigner transformation~\cite{wigner1932quantum}, $f_N(\mathbf{q},\mathbf{p})$ is obtained. The equation of motion of $f_N(\mathbf{q}, \mathbf{p})$ reads as,
\begin{eqnarray} \label{eq-fnt-append}
&&\frac{\partial f_N(\mathbf{q},\mathbf{p},t)}{\partial t} = \frac{1}{\left( \pi 
		\hbar \right)^N} \int  dy_1 \cdots dy_N e^{-2i \mathbf{ p } \cdot \mathbf{ y } / 
		\hbar}\\\nonumber
&& \left\{ \frac{\partial g_N^*(\mathbf{q-y})}{\partial t}g_N(\mathbf{q+y})  
		+ g_N^*(\mathbf{q-y})\frac{\partial g_N(\mathbf{q+y})}{\partial t} \right\}.
\end{eqnarray}
where $\mathbf{p}$ is the conjugate momentum of $\mathbf{q}$. 

In the derivations of Eq.~(\ref{eq-fnt-append}), the $\partial g_N^*/\partial t$ and  
$\partial g_N/\partial t$ are replaced with the right hand side of Eq.~(\ref{eq-g-append}) 
and we have,
%
%
\begin{eqnarray}\label{eq-fnt-expand}
&&\frac{\partial f_N(\mathbf{q}, \mathbf{p},t)}{\partial t} = \frac{1}{\left( \pi 
	\hbar \right)^N} \int  dy_1 \cdots dy_N e^{-2i \mathbf{ p } \cdot \mathbf{ y }/\hbar}\\\nonumber
&& \bigg\{\sum\limits_{kl} \bigg[ -\frac{i\hbar }{2}\frac{\partial } {\partial 
		(q_{k}-y_k)} \bigg ({B}_{kl} \left( \mathbf{q-y} \right) \frac {\partial {{g}_{N}^{*}}\left(\mathbf{q-y} \right)} {\partial (q_l-y_l)}\bigg )
	g_{N} ( \mathbf{q+y} ) \\ \nonumber
&&	+ \frac{i\hbar }{2} g^*_{N} ( \mathbf{q-y} )\frac{\partial } {\partial 
	(q_{k}+y_k)}\bigg ( {B}_{kl} \left( \mathbf{q+y} \right)  \frac {\partial {{g}_{N}}\left(\mathbf{q+y} \right)} {\partial (q_l+y_l)}\bigg) \bigg ]\\\nonumber
&& + \frac{i}{\hbar}\left ( V(\mathbf{q-y})-V(\mathbf{q+y})\right )g^*_{N} ( \mathbf{q-y} )g_{N} ( \mathbf{q+y} )
\bigg\}.
\end{eqnarray}
One should note that the kinetic energy term contains the collective coordinate $\mathbf{ 
q}$ dependence of inertia $B_{kl}$, which is different than the N-body system with fixed 
particle mass.

In coordinate space, the kinetic energy terms in Eq.~(\ref{eq-fnt-expand}) are as follows,
\begin{eqnarray} \label{eq220921182431}
	&&\mathcal M_{kl}=- \frac{i \hbar}{2}\frac{1}{(\pi\hbar)^N} \int  \text{d} y_1 \cdots \text{d} y_N  e^{- 2i \mathbf{p}\cdot\mathbf{y} / \hbar } \\\nonumber
	&&\bigg[\frac{\partial  }{\partial y_k}\bigg (B_{kl} ( \mathbf{ q-y}) \frac{\partial 
		g_N^* ( \mathbf{q} - \mathbf{y} )}{ \partial y_l }\bigg ) g_N ( \mathbf{ q } + \mathbf{y} )\\\nonumber
	&&-	g_N^* ( \mathbf{q} - \mathbf{y} )\frac{\partial  }{\partial y_k}\bigg(B_{kl} ( \mathbf{ q+y}) \frac{\partial 
	g_N ( \mathbf{ q } + \mathbf{y} ) }{ \partial y_l }\bigg)
	\bigg]\\\nonumber
	&&=\mathcal M_{kl,1}+\mathcal M_{kl,2}
	.
\end{eqnarray}

Since the $g_N$ and $g_N^*$ depend on the integration variable $y_k$, one can replace the 
differentiations with respect to $q_k+ y_k$ by differentiations with respect to $y_k$ in 
derivations. By doing one partial integration with respect to $y_k$, one has
\begin{eqnarray}\label{keterm}
&&\mathcal{M}_{kl,1}	= \\\nonumber
&& - \frac{i\hbar}{2} \frac{1}{(\pi\hbar)^N} B_{kl} ( \mathbf{q-y} ) \frac{ \partial g_N^* ( \mathbf{q-y} ) }{ \partial y_l } g_N ( \mathbf{q} + \mathbf{y} ) e^{ -2i \mathbf{p}\cdot\mathbf{y} / \hbar } 
	\bigg |_{- \infty}^{+\infty} \\\nonumber
&&+\frac{i\hbar}{2} \frac{1}{(\pi\hbar)^N}\int \text{d} y_1 \cdots \text{d} y_N B_{kl} ( \mathbf{q-y} )\frac{ \partial g_N^* ( \mathbf{q-y} )}{\partial y_l }\\\nonumber
&& \times \frac{\partial}{\partial y_k}\left (  g_N ( \mathbf{ q + y} ) e^{ -2i \mathbf{p}\cdot \mathbf{y} / \hbar } \right ) \\\nonumber
\end{eqnarray}
The first term in Eq.~(\ref{keterm}) vanishes due to the boundary condition at infinity, and second term becomes
\begin{eqnarray}\label{keterm-simp}
	\mathcal{M}_{kl,1}&=& \frac{i \hbar}{2} \frac{1}{(\pi\hbar)^N}\int \text{d} y_1 \cdots \text{d} y_N \\\nonumber
	&&\times B_{kl} ( \mathbf{q}- 
	\mathbf{y} )\frac{\partial g_N^* ( \mathbf{q-y } )}{ \partial y_l }  \frac{\partial}{\partial y_k} \left (  g_N ( \mathbf{ q+y} ) e^{ -2i \mathbf{p}\cdot\mathbf{y}/ \hbar }\right )\\\nonumber
&=& \frac{i \hbar}{2} \frac{1}{(\pi\hbar)^N}\int\text{d} y_1 \cdots \text{d} y_N s  \\\nonumber
	&&\times B_{kl} ( \mathbf{q}-\mathbf{y} )\frac{\partial g_N^* ( \mathbf{q-y } )}{ \partial y_l } \bigg [ \frac{\partial g_N ( \mathbf{ q+y} )}{\partial y_k}  e^{ -2i \mathbf{p}\cdot\mathbf{y}/ \hbar }\\\nonumber
	&&+  g_N ( \mathbf{ q+y} ) e^{ -2i \mathbf{p}\cdot\mathbf{y}/ \hbar }(-\frac{2ip_k}{\hbar})\bigg ]\\
&=&\frac{i \hbar}{2} B_{kl}(\mathbf{q-y^*_{(1a)}})\frac{1}{(\pi\hbar)^N}\int\text{d} y_1 \cdots \text{d} y_N \\\nonumber  
	&&\frac{\partial g_N^* ( \mathbf{q-y } )}{ \partial y_l } \frac{\partial  g_N(\mathbf{q+y})}{\partial y_k}  e^{- 2i \mathbf{p}\cdot\mathbf{y}/ \hbar } \\\nonumber
	&&+\frac{i \hbar}{2} B_{kl}(\mathbf{q-y^*_{(1b)}})\frac{1}{(\pi\hbar)^N}\int\text{d} y_1 \cdots \text{d} y_N \\\nonumber
	&&\frac{\partial g_N^* ( \mathbf{ q -y } ) }{ \partial y_l } g_N ( \mathbf{ q +y} ) e^{- 2i \mathbf{p}\cdot\mathbf{y}/ \hbar } (-\frac{2ip_k}{\hbar}).
\end{eqnarray} 
In above derivations, the $B_{kl}(\mathbf{q- y})$ is moved out by assuming the following relationship, i.e., 
\begin{eqnarray}\label{Int-Approx-App}
	&&\int B_{kl}(\mathbf{q-y}) O(\mathbf{q},\partial/\partial \mathbf{q})dy_1\cdots dy_{N}=\\\nonumber
	&&B_{kl}(\mathbf{q-y^*_{(I)}})\int  O(\mathbf{q},\partial/\partial \mathbf{q})dy_1\cdots dy_{N}.
\end{eqnarray}

Similarly, the $\mathcal M_{kl,2}$ becomes
 \begin{eqnarray}\label{keterm-m2}
 	\mathcal{M}_{kl,2}=-\frac{i \hbar}{2} B_{kl}(\mathbf{q+y^*_{(2a)}})\frac{1}{(\pi\hbar)^N}\int\text{d} y_1 \cdots \text{d} y_N \\\nonumber  
	\frac{\partial g_N^* ( \mathbf{q-y } )}{ \partial y_k } \frac{\partial  g_N(\mathbf{q+y})}{\partial y_l}  e^{- 2i \mathbf{p}\cdot\mathbf{y}/ \hbar } \\\nonumber
	-\frac{i \hbar}{2} B_{kl}(\mathbf{q+y^*_{(2b)}})\frac{1}{(\pi\hbar)^N}\int\text{d} y_1 \cdots \text{d} y_N \\\nonumber
	g_N^* ( \mathbf{ q -y} )\frac{\partial g_N ( \mathbf{ q +y } ) }{ \partial y_l }  e^{- 2i \mathbf{p}\cdot\mathbf{y}/ \hbar } (-\frac{2ip_k}{\hbar})
 \end{eqnarray}
%

%
By defining $\bar{\mathcal{B}}_{kl}^{(a)}$, $\bar{\mathcal{B}}_{kl}^{(b)}$, $\delta \mathcal{B}_{kl}^{(a)}$ and $\delta\mathcal{B}_{kl}^{(b)}$ as,
\begin{align*}
\bar{\mathcal{B}}_{kl}^{(a)} &= \bigg( B_{kl}(\mathbf{q- y^*_{(1a)}})+B_{kl}(\mathbf{q+ y^*_{(2a)}})\bigg )/2,\\
\bar{\mathcal{B}}_{kl}^{(b)} &= \bigg( B_{kl}(\mathbf{q- y^*_{(1b)}})+B_{kl}(\mathbf{q+ y^*_{(2b)}})\bigg )/2,\\
\delta \mathcal{B}_{kl}^{(a)} &= \bigg( B_{kl}(\mathbf{q- y^*_{(1a)}})-B_{kl}(\mathbf{q+ y^*_{(2a)}})\bigg ),\\
\delta \mathcal{B}_{kl}^{(b)} &= \bigg( B_{kl}(\mathbf{q- y^*_{(2a)}})-B_{kl}(\mathbf{q+ y^*_{(2b)}})\bigg ),  
\end{align*}
and assuming $\delta \mathcal{B}_{kl}^{(a)}\approx0$ and $\delta\mathcal{B}_{kl}^{(b)}\approx 0$, $\mathcal M_{kl}$ becomes
\begin{eqnarray}\label{keterm-total}
	\mathcal M_{kl}&=&\mathcal M_{kl,1}+\mathcal M_{kl,2}\\\nonumber
	&=& -\frac{i \hbar}{2}\bar{\mathcal {B}}^{(a)}_{kl} \frac{1}{(\pi\hbar)^N}\int\text{d} y_1 \cdots \text{d} y_N e^{- 2i \mathbf{p}\cdot\mathbf{y}/ \hbar } \\\nonumber
	&& \bigg [\frac{\partial g_N^* ( \mathbf{q-y } )}{ \partial q_l } \frac{\partial  g_N(\mathbf{q+y})}{\partial q_k} -\frac{\partial g_N^* ( \mathbf{q-y } )}{ \partial q_k } \frac{\partial  g_N(\mathbf{q+y})}{\partial q_l} \bigg ] \\\nonumber
    && -\frac{i \hbar}{2}\bar{\mathcal {B}}^{(b)}_{kl}  \frac{1}{(\pi\hbar)^N} \int\text{d} y_1 \cdots \text{d} y_N e^{-2i \mathbf{p}\cdot\mathbf{y}/ \hbar }\\\nonumber
    &&\bigg [ \frac{\partial g^*_N ( \mathbf{ q } - \mathbf{ y } ) }{ \partial q_l } g ( \mathbf{ q } + \mathbf{y} )-  g^*_N ( \mathbf{ q } - \mathbf{y} ) \frac{ \partial g_N ( \mathbf{ q } + \mathbf{ y } ) }{ \partial q_l } \bigg ]\\\nonumber
    &&(-\frac{2ip_k}{\hbar}).
\end{eqnarray}
In Eq.(\ref{keterm-total}), an identical relationship between differential with respect to $y_k(y_l)$ and with respect to $q_k(q_l)$ is used.

Due to the symmetric property of $\bar{\mathcal {B}}^{(a)}_{kl}$ and summation of $\sum_{kl}$ in Eq.(\ref{eq-fnt-expand}), the contributions from first term in Eq.(\ref{keterm-total}) vanishes. Thus, the kinetic part can be written as,
\begin{equation} \label{eq221010165740}
	\mathcal M_{kl}= -p_k\bar{\mathcal{B}}_{kl}^{(b)} \frac{\partial f_N}{\partial q_l}
\end{equation}

For the potential part in Eq.~(\ref{eq-fnt-expand}),  a Taylor series with respect to 
$\mathbf q$ is performed with potential fields $ V_N \left( \mathbf{ q } + \mathbf{ y } 
\right) $ and $ V_N \left( \mathbf{ q } - \mathbf{ y }  \right) $. 
\begin{eqnarray}
    \mathcal N &=& \frac{1}{\left( \pi\hbar \right)^N} \int  dy_1 \cdots dy_N e^{-2i \mathbf{ p } \cdot \mathbf{ y } / \hbar }\\\nonumber
&&\times\frac{i}{\hbar}\left[ V(\mathbf{q-y})-V(\mathbf{q+y})\right ]g^*_{N} ( \mathbf{q-y} )g_{N} ( \mathbf{q+y} )	\\\nonumber
&=&\frac{1}{\left( \pi	\hbar \right)^N} \int  dy_1 \cdots dy_N e^{-2i \mathbf{ p } \cdot \mathbf{ y }}\\\nonumber
&&\times\frac{ 2i }{\hbar} \sum\limits_{\lambda} \frac{1}{\lambda_1 ! \cdots \lambda_N !} \frac{ 
		\partial^{\lambda_1+ \cdots \lambda_N} V_N \left( \mathbf{q} \right) } {\partial 
		q_1^{\lambda_1} \cdots \partial q_N^{\lambda_N}} \\\nonumber
&&	y_1^{\lambda_1} \cdots y_N^{ \lambda_N} g_N^* \left( \mathbf{ q - y } \right) g_N \left( \mathbf{ q +y } \right)\\\nonumber
&=&\frac{2i}{\hbar}\sum\limits_{\lambda} (\frac{\hbar}{2i})^{\lambda_1 + \cdots + \lambda_N}\frac{1}{\lambda_1 ! \cdots \lambda_N !}\\\nonumber 
&&\times\frac{ 
		\partial^{\lambda_1 + \cdots + \lambda_N} V_N \left( \mathbf{q} \right) } {\partial 
		q_1^{\lambda_1} \cdots \partial q_N^{\lambda_N}} 
	\frac{ \partial^{\lambda_1 + \cdots + \lambda_N} f_N  } {\partial 
		p_1^{\lambda_1} \cdots \partial p_N^{\lambda_N}}
\end{eqnarray}

Finally, Eq.~\eqref{eq-fnt-expand} could be written as

\begin{eqnarray}\label{eq007}
\frac{\partial f_N(\mathbf{q},\mathbf{p},t)}{\partial t} &=& -\sum_{kl} p_k \bar{\mathcal {B}}_{kl}^{(b)} ( \mathbf{q} ) \frac{\partial f_N}{\partial q_l} \\\nonumber
&& +\sum\limits_{\lambda}(\frac{\hbar}{2i})^{\lambda_1 + \cdots + \lambda_N-1} \frac{1}{\lambda_1 ! \cdots \lambda_N !} \\\nonumber
&&\times\frac{ 
		\partial^{\lambda_1 + \cdots + \lambda_N} V_N \left( \mathbf{q} \right) } {\partial 
		q_1^{\lambda_1} \cdots \partial q_N^{\lambda_N}} 
	\frac{ \partial^{\lambda_1 + \cdots + \lambda_N} f_N  } {\partial 
		p_1^{\lambda_1} \cdots \partial p_N^{\lambda_N}}\\\nonumber
 &=& -\sum_{kl} p_k \bar{\mathcal {B}}_{kl}^{(b)} ( \mathbf{q} ) \frac{\partial f_N}{\partial q_l}+\sum_l^N \frac{\partial V}{\partial q_l}\cdot\frac{\partial f_N}{\partial p_l}\\\nonumber
&&+\sum\limits_{\lambda}(\frac{\hbar}{2i})^{\lambda_1 + \cdots + \lambda_N-1} \frac{1}{\lambda_1 ! \cdots \lambda_N !} \\\nonumber
&&\times\frac{ 
		\partial^{\lambda_1 + \cdots + \lambda_N} V_N \left( \mathbf{q} \right) } {\partial 
		q_1^{\lambda_1} \cdots \partial q_N^{\lambda_N}} 
	\frac{ \partial^{\lambda_1 + \cdots + \lambda_N} f_N  } {\partial 
		p_1^{\lambda_1} \cdots \partial p_N^{\lambda_N}}.		.
\end{eqnarray}
where all $\lambda_1, \cdots, \lambda_N $ are non-negative integer values and $\lambda_1 +
\cdots + \lambda_N $ is an odd number. This is a general form of transport equation for TDGCM+GOA.
When the potential is calculated from the two-body interaction, i.e.,
\begin{equation}
	V(q_1, q_2, \cdots, q_N)=\sum_{i\le j} V(q_i,q_j), 
\end{equation}
the equation is further simplified as,
\begin{equation}\label{eq010}
\frac{\partial f_N}{\partial t} = -\sum_{kl}  \bar{\mathcal {B}}_{kl}^{(b)} ( \mathbf{q} ) p_k\frac{\partial f_N}{\partial q_l} 
+\sum_{ k\le m } \frac{\partial V_{km}}{\partial q_k} \frac{ \partial f_N } { \partial p_k }.
\end{equation}

\section{Time evolution of $s$-body phase space density distribution}\label{app-fst}

The $s$-body phase space density is defined as,
\begin{eqnarray}\label{fs-app}
&&	f_s(q_1, \cdots, q_s, p_1, \cdots, p_s)\\\nonumber
&&=\frac{1}{\Omega^{N-s}}\int f_N(q_1, \cdots, q_N, p_1, \cdots, p_N)d\Gamma_{s+1}\cdots d\Gamma_N,\\\nonumber
&&	 d\Gamma_i=dq_idp_i.
\end{eqnarray}
When the potential is calculated from the two-body interaction, the time-dependent probability distribution $f_s$ is obtained by the similar strategy of the derivation of the Bogoliubov-Born-Green-Kirkood-Yvon (BBGKY) hierarchy, i.e.,
\begin{eqnarray}\label{fs-t}
	&&	\frac{\partial f_s(q_1, \cdots, q_s, p_1, \cdots, p_s)}{\partial t} \nonumber \\
	&&=\frac{1}{\Omega^{N-s}}\int \text{d} \Gamma_{s+1}\cdots \text{d} \Gamma_N \frac{\partial f_N(q_1, \cdots, q_N, p_1, \cdots, p_N)}{\partial t} \nonumber \\
	&&=\frac{1}{\Omega^{N-s}}\int \text{d} \Gamma_{s+1}\cdots \text{d} \Gamma_N \bigg[-\sum_{kl}  \bar{\mathcal {B}}_{kl}^{(b)} ( \mathbf{q} ) p_k\frac{\partial f_N}{\partial q_l} \nonumber \\
	&& \qquad + \sum_{ k\le m } \frac{\partial V_{km}}{\partial q_k} \frac{\partial f_N } { \partial p_k } \bigg].
\end{eqnarray}
Different than the transport equation for fixed-mass many-particle system, the inertia $B_{kl}(\mathbf{q})$ depends on the coordinate and it causes the equation much more complexity.

To avoid the difficulty caused by $\bar{\mathcal{B}}_{kl}(\mathbf{q})$, we move the $\bar{\mathcal{B}}_{kl}$ out from the 
integration by using the equivalent of the integration, i.e., 
\begin{eqnarray}
	&&\int \bar{\mathcal{B}}_{kl}(\mathbf{q}) O(\mathbf{q},\mathbf{p})d\Gamma_{s+1}\cdots d\Gamma_{N}=\\\nonumber
	&&\bar{\mathcal{B}}_{kl}(\mathbf{q_s}, q_{s+1}^*,\cdots, q_{N}^*)\int  O(\mathbf{q},\mathbf{p})d\Gamma_{s+1}\cdots d\Gamma_{N}.
\end{eqnarray}
Consequently, the derivation will be simplified. 

For the first term in r.h.s of Eq.~(\ref{fs-t}), we
perform one partial integration with respect to $q_l$ and the result is
\begin{align}\label{fs-t-simp}
I_1=&-\frac{1}{\Omega^{N-s}}\int\bigg[\sum_{kl}  \bar{\mathcal {B}}_{kl}^{(b)} ( \mathbf{q} ) p_k\frac{\partial f_N}{\partial q_l}\bigg]d\Gamma_{s+1}\cdots d\Gamma_N \nonumber \\
=&-\sum_{kl} \bar{\mathcal {B}}_{kl}^{(b)} (\mathbf{q_s},q_{s+1}^*,\dots, q_{N}^*) \nonumber \\
& \qquad \qquad \times\frac{1}{\Omega^{N-s}}\int\bigg[   p_k\frac{\partial f_N}{\partial q_l}\bigg]d\Gamma_{s+1}\cdots d\Gamma_N \nonumber \\
=&-\sum_{k=1}^s\sum_{l=1}^s \bar{\mathcal {B}}_{kl}^{(b)} (\mathbf{q_s},q_{s+1}^*,\dots, q_{N}^*) p_k\frac{\partial f_s}{\partial q_l} \nonumber \\
& \qquad \qquad -\sum_{k=s+1}^N\sum_{l=1}^s \bar{\mathcal {B}}_{kl}^{(b)} (\mathbf{q_s},q_{s+1}^*,\dots, q_{N}^*) \nonumber \\
& \qquad \qquad \times\frac{1}{\Omega^{N-s}}\int  p_k\frac{\partial f_N}{\partial q_l}d\Gamma_{s+1}\cdots d\Gamma_N \nonumber \\
=&-\sum_{k=1}^s\sum_{l=1}^s \bar{\mathcal {B}}_{kl}^{(b)} (\mathbf{q_s},q_{s+1}^*,\dots, q_{N}^*) p_k\frac{\partial f_s}{\partial q_l}-\delta I_1
\end{align}
In above derivation, we use the term of $\sum_{l=s+1}^N \bar{\mathcal {B}}_{kl}^{(b)} (\mathbf{q}) p_k 
f_N|_{q_l\to -\infty}^{q_l\to\infty}=0$. This is because the finite values of $f_N$, which 
means $f_N(\mathbf q)=0$ at $q\to{\pm\infty}$. The $\delta I_1$ is defined as,
\begin{eqnarray}
\delta I_1&=&\sum_{k=s+1}^N\sum_{l=1}^s \bar{\mathcal {B}}_{kl}^{(b)} (\mathbf{q_s},q_{s+1}^*,\dots, q_{N}^*)\\\nonumber
&& \qquad \qquad \cdot \frac{1}{\Omega^{N-s}}\int  p_k\frac{\partial f_N}{\partial q_l}d\Gamma_{s+1}\cdots d\Gamma_N\\\nonumber
&=&\sum_{k=s+1}^N\sum_{l=1}^s \bar{\mathcal {B}}_{kl}^{(b)} (\mathbf{q_s},q_{s+1}^*,\dots, q_{N}^*) p_k^*\frac{\partial f_s}{\partial q_l}
\end{eqnarray}
which is connected to the $N$-body density distribution $f_N$, and reflects how the 
momentum field evolve with the coordinates.

The second term in Eq.~(\ref{fs-t}) reads,
\begin{eqnarray}
	I_3&=&\frac{1}{\Omega^{N-s}}\int \sum_{ k\le m } \frac{\partial V_{km}}{\partial q_k} \frac{\partial f_N } { \partial p_k }d\Gamma_{s+1}\cdots d\Gamma_N\\\nonumber
	&=& \frac{1}{\Omega^{N-s}}\int \sum_{1\le k\le m\le s } \frac{\partial V_{km}}{\partial q_k} \frac{\partial f_N } { \partial p_k }d\Gamma_{s+1}\cdots d\Gamma_N\\\nonumber
	&+&(N-s)\frac{1}{\Omega^{N-s}}\int \sum_{k=1}^l \left(\frac{\partial V_{k,s+1}}{\partial q_k}\right)\left(\frac{\partial f_N}{\partial p_k}\right)d\Gamma_{s+1}\cdots d\Gamma_{N}\\\nonumber
	&=&\sum_{1\le k\le m\le s } \frac{\partial V_{km}}{\partial q_k} \frac{\partial f_s } { \partial p_k }+\frac{N-s}{\Omega}\int \sum_{k=1}^s \left(\frac{\partial V_{k,s+1}}{\partial q_k}\right)\\\nonumber
	&\times&\left(\frac{\partial f_{s+1}}{\partial p_k}\right)d\Gamma_{s+1}
\end{eqnarray}

Finally, the time evolution of $f_s$ is,
\begin{eqnarray}\label{fst-final-append}
	\frac{\partial f_s}{\partial t}&=&-\sum_{k=1}^s\sum_{l=1}^s \bar{\mathcal {B}}_{kl}^{(b)} (\mathbf{q_s},q_{s+1}^*,\dots, q_{N}^*) p_k\frac{\partial f_s}{\partial q_l}\\\nonumber
	&-&\sum_{k=s+1}^N\sum_{l=1}^s \bar{\mathcal {B}}_{kl}^{(b)} (\mathbf{q_s},q_{s+1}^*,\dots, q_{N}^*) p_k^*\frac{\partial f_s}{\partial q_l}\\\nonumber
	&-&\sum_{1\le k\le m\le s } \frac{\partial V_{km}}{\partial q_k} \frac{\partial f_s } { \partial p_k }+\frac{N-s}{\Omega}\int \sum_{k=1}^s \left(\frac{\partial V_{k,s+1}}{\partial q_k}\right)\\\nonumber
	&\times&\left(\frac{\partial f_{s+1}}{\partial p_k}\right)d\Gamma_{s+1}
\end{eqnarray}
\bibliography{apssamp}
\end{document}